\documentclass[floatfix,aps,pre,twocolumn,superscriptaddress,showpacs]{revtex4}
\usepackage{epsfig,amsmath,amssymb,graphicx,color,calc,bm,bbold,ulem}

\let\a\alpha  \let\d\delta \let\l\lambda \let\r\rho \let\th\theta
\let\Th\Theta \let\S\Sigma

\def\io{\infty}
\def\hr{\hat\r}
\def\hth{\hat\theta}

\def\to{\rightarrow}
\def\la{\left\langle}
\def\ra{\right\rangle}

\newcommand{\beq}{\begin{equation}}
\newcommand{\eeq}{\end{equation}}
\newcommand{\ba}{\begin{align}}
\newcommand{\ea}{\end{align}}

\begin{document}

\title{Equilibrium dynamics of the Dean-Kawasaki equation: MCT and beyond}

\author{Bongsoo Kim}
\affiliation{Department of Physics and Institute for Soft and Bio Matter Science, Changwon National University, Changwon 641-773, Korea}

\author{Kyozi Kawasaki}
\affiliation{Department of Physics, Faculty of Science, Kyushu University, 
Fukuoka 812-8581, Japan}

\author{Hugo Jacquin}
\affiliation{Laboratoire Mati\`ere et Syst\`emes Complexes, UMR7057 CNRS/Paris 7, Universit\'e Denis Diderot, 10 rue Alice Domon et L\'eonie Duquet, 75205 Paris cedex 13, France}

\affiliation{Laboratoire de Physique, UMR5672 CNRS/ENS Lyon, \'Ecole Normale Sup\'erieure de Lyon, 46 all\'e d'Italie, 69364 Lyon cedex 07, France}

\author{Fr\'ed\'eric van Wijland}
\affiliation{Laboratoire Mati\`ere et Syst\`emes Complexes, UMR7057 CNRS/Paris 7, Universit\'e Denis Diderot, 10 rue Alice Domon et L\'eonie Duquet, 75205 Paris cedex 13, France}

\date{\today}

\begin{abstract}
We extend a previously proposed field-theoretic self-consistent perturbation approach
for the equilibrium dynamics of the Dean-Kawasaki equation presented in [J. Stat. Mech. {\bf 2008} P02004]. 
By taking terms missing in the latter analysis into account we arrive at a set of three new
equations for correlation functions of the system. 
These correlations involve the density and its logarithm as local observables. 
Our new one-loop equations, 
which must carefully deal with the noninteracting Brownian gas theory, are more general than the historic Mode-Coupling one in that a further and 
well-defined approximation leads back to the original mode-coupling equation for the density correlations alone. However, without performing any further 
approximation step, our set of three equations does  
not feature any ergodic-non ergodic transition, as opposed to the historical mode-coupling approach.
\end{abstract}



\maketitle

\tableofcontents

\newpage
\section{Introduction}
\label{intro}

Supercooled liquids approaching the glass transition exhibit fascinating dynamic phenomena 
such as tremendous slowing down and kinetic heterogeneity \cite{dasbook,BB11}. 
These dynamic properties are intimately connected to the nature of the transition from liquid to glass. 
First-principle theoretical understanding of the dynamics of supercooled liquids and the nature of their 
glass formation still remains one of the greatest challenges in condensed matter science.

Applications of the mode coupling theory (MCT) to supercooled liquids \cite{goetze} started in 
1984, and there are still certain aspects of it that remain controversial. The best known result is 
the derivation of the standard MCT (SMCT) equation for the dynamical structure factor. 
This equation predicts the existence of a sharp kinetic transition between 
fluid-like ergodic and glass-like non-ergodic states. However, it is a widespread belief that 
this sharp transition to full dynamical arrest is spurious, and that a sharp cross-over must instead take place \cite{clothes,MMR06}. 
The latter  could result from thermally activated processes left out from the SMCT equation. The absence of a sharp transition
described by SMCT is now well established numerically \cite{BEPBCPS09,EBPPSBC09}, 
but the theoretical extensions of SMCT devised in order to take the so-called ``activated events" \cite{DM86,GS87} into account 
is even more controversial than SMCT itself \cite{CR06,DM09}.
These extensions attempt to include thermally activated 
processes which are expected to round off the sharp ergodic-nonergodic (ENE) transitions, 
by including the current variables in the description.
But if the current variables relax in time much faster than the density variables 
as in deeply supercooled liquids, the former variables can adiabatically be eliminated to result in 
a closed stochastic equation for the density modes.
This is called the dynamical density functional theory where 
thermally activated processes are also taken into account \cite{MK90,KK94}.
This is evident in appearance of the second functional derivative with respect to the density 
variable in the stochastic equation for the probability density functional (see Eq.~(\ref{eq2.5}) below). 
This guarantees the approach to equilibrium of the density probability distribution functional as the time goes to infinity. 
In any case, the search for a proper theoretical description that 
incorporates density fluctuations alone is needed, at least in order to treat Brownian dynamics, in which no energy or momenta currents exist.

With these issues in the background, 
in a recent work \cite{KK08} (referred to as [KK08] throughout the present work), two of the 
present authors investigated the dynamical density functional equation for colloidal suspension, 
that we denote here by the Dean-Kawasaki (DK) equation \cite{KK94,De96}. 
In one loop order [KK08] claimed derivation of SMCT which predicts a sharp ENE transition.
[KK08] anticipated at that time that smearing effects for this transition should come from higher 
loop, which, however, is not the case. As we first show in this paper, there is a missing term in 
[KK08], and recovering it gives rise to a new set of equations that bear on several correlation functions instead of only one. 
We will show that  
these equations with one-loop self energies 
do not support the existence of a sharp ENE transition. 
We then demonstrate a striking property of these equations: that SMCT can be recovered from them by a further, well defined (but again not justified) approximation.

\section{The Dean-Kawasaki equation}
\label{DK}

The DK equation is the following stochastic time evolution equation for the  density field $\r({\bf r},t)$  
\beq
\partial_t \r({\bf r},t) \! = \! \nabla \cdot \left( \r({\bf r},t) \nabla \frac{\d F[\r]}
{\d \r({\bf r},t)} \right) +  \nabla \cdot \left( \frac{}{} \!\! \sqrt{ 2 T \r ({\bf r},t)} 
{\boldsymbol \eta} ({\bf r},t) \right), 
\label{eq2.1}
\eeq
where $T$ is the temperature of the system (we set the Boltzmann constant $k_B$ unity throughout).
In Eq.~(\ref{eq2.1}),  the Gaussian thermal noise $\eta_{\a}({\bf r},t)$ has zero mean and unit variance 
\beq
\la \eta_{\a}({\bf r},t) \eta_{\beta} ({\bf r}',t') \ra = \d_{\a \beta} \d({\bf r}-{\bf r}') \d(t-t') \ .
 \label{eq2.2}
\eeq
The free energy density functional $F[\r]$ reads:
\beq\begin{split}
F[\r] &=  F_{\rm{id}}[\r]+ F_{\rm{int}} [\r] \ ,  \\
F_{\rm{id}}[\r] &= T \int d{\bf r} \,  \r ({\bf r}) \left[\ln \left(\frac{\r ({\bf r})}{\r_0}\right)-1 \right] \ ,  \\
F_{\rm{int}} [\r] &= \frac{1}{2} \int  d {\bf r} \int  d {\bf r}' \,  U(|{\bf r}-{\bf r}'|) \d \r ({\bf r}) \,  \d \r ({\bf r}') \ , 
\end{split}
\label{eq2.3}
\eeq
where $\d \r ({\bf r},t) \equiv \r({\bf r},t)-\r_0$
is the density fluctuation around the equilibrium density $\r_0$.
In Eq.~(\ref{eq2.3}) $F_{\rm{id}}[\r]$ is the entropic contribution to the free-energy, and $F_{\rm{int}}[\r]$ 
the interaction one with $U(|{\bf r}-{\bf r}'|)$ representing the particle interaction.
This equation, obtained by Dean \cite{De96} is the {\it exact} evolution equation for the microscopic density of a 
collection of interacting Brownian particles under overdamped dynamics. It thus contains all the complexity of the systems 
that undergo a glass transition in numerical simulations, or colloidal experiments. 
Surprisingly, Eqs.~(\ref{eq2.1})--(\ref{eq2.3}) were proposed by one of the present authors \cite{KK94} as 
a mesoscopic kinetic equation for the coarse-grained density with $U(r)=-Tc(r)$ ($c(r)$ being the direct correlation function of the liquid). 
In particular, Eq.~(\ref{eq2.1}) was obtained via adiabatic elimination of the much faster-decaying momentum field 
in the fluctuating hydrodynamic equations of dense liquids \cite{DM86}. 
We wish to emphasize that in this work, we do {\it not} use this substitution at any place, and the appearance of the 
renormalized static correlation function $c(r)$ occurs naturally, due to the compatibility between equilibrium and dynamic correlations
in the formalism we use, and the correct treatment of non-perturbative sum rules enforced by this compatibility.

A prominent feature of the DK equation in Eq.~(\ref{eq2.1}-\ref{eq2.3})
 is that the diffusion part has an extra factor of density, and hence
the corresponding thermal noise has to be of multiplicative form in order for the
density modes to relax towards their expected Gibbs distribution. The multiplicative nature of the noise
greatly complicates the theoretical treatment of the dynamics of the system.
It is also crucial to give a diffusion equation for the density fluctuations in the absence of interaction due to the 
non-polynomial density dependence of $F_{\rm{id}}[\r]$;
\beq
\frac{\d F_{\rm{id}}}{\d \r({\bf r})} = T \ln \frac{\r({\bf r})}{\r_0} \text{ hence }
\nabla \cdot \left( \r \nabla \frac{\d  F_{id}}{\d \r} \right)  = T \nabla^2 \r \ .
\label{eq2.4}
\eeq
This feature is physically expected (and analytically proved in \cite{VCCK08}) for the non-interacting Brownian particles.
The second element is essential for the system 
to evolve toward the equilibrium stationary state due to the presence of the extra factor of density in the diffusion part.
Specifically, the  Fokker-Planck (FP) equation corresponding to Eqs.~(\ref{eq2.1})~and~(\ref{eq2.2}) can be written as
\begin{align}
& \frac{\partial P( \{\r\},t)}{\partial t} = \label{eq2.5} \\
& -\int d{\bf r} \,\, \frac{\d}{\d \r({\bf r})}
\nabla \cdot \r ({\bf r}) \nabla \left[ T \frac{\d}{\d \r({\bf r})}
+\frac{\d F[\r]}{\d \r({\bf r})}  \right] P( \{\r\},t) \ , \nonumber
\end{align}
where $P( \{\r\},t) $ is the probability distribution of the density configuration $\{\r({\bf r})\}$ at time $t$.
The equilibrium distribution $P_{\rm{eq}} [\r] \propto \exp \big(-F [\r] /T \big)$ 
is a stationary solution of the FP equation Eq.~(\ref{eq2.5}).
As stated in the Introduction, in view of the presence of the second functional derivative with respect to the density variable in 
Eq.~(\ref{eq2.5}), which comes from the multiplicative
thermal noise in (\ref{eq2.1}),
thermally activated processes are included in the equation. 
Of course this is a weak analogy to Kramer's theory for the escape across a one-dimensional potential energy barrier,
where the second derivative of the free-energy represents the local curvature, which governs the escape rate.
But at present we do not know an analytic way of handling the above Fokker-Planck-type equation to produce activated processes. 
We here just quote Langer's theory \cite{La69} where the Fokker-Planck-type equation 
is used to calculate nucleation rate of first order transition using instanton-type non-perturbative calculation,  which is therefore outside
any renormalized perturbation theory (RPT) including the present one.

A systematic way of analyzing the DK equation is to perform a RPT (i.e. the loop expansion) 
on the dynamic action $S[\r,\hr] $ obtained from the original  DK equation, which enables one to treat dynamics of the correlation and response functions on equal footing. 
Via the Martin-Siggia-Rose-Janssen-De Dominicis method 
\cite{MSR73,Ja79,Do76}
one can obtain  the dynamic action 
 ${\cal S} [\r, \hat \r ]$ for (\ref{eq2.1}) and (\ref{eq2.2})
\beq
{\cal S} [ \r, \hat \r ] = \int_{{\bf r},t} 
 \left\{ i\hat \r \left[ \partial_t \r -  \nabla \cdot \left( \r \nabla 
\frac{\d F[\r]}{\d \r} \right)  \right] -T \r ( \nabla \hr)^2 \right\} \ ,
 \label{eq2.6}
\eeq
where $\int_{{\bf r},t} \equiv \int d{\bf r} \int dt $,
the auxiliary field $\hr$ is a real field,  and the last cubic term involving the quadratic $\hr $ comes from the average over the multiplicative thermal noise $\eta$. 
The dynamic action of this form with the Ramakrishnan-Yousouff (RY) free energy functional, i.e., the free energy functional in Eq.~(\ref{eq2.3}) with $U(r)=-Tc(r)$,  was first written down in \cite{KK97}. 
In deriving Eq.~(\ref{eq2.6}), employing  the It\^{o} calculus  makes the Jacobian of the transformation constant \cite{Ja79,Ja92}. 
In principle, studying time-reversibility in such dynamic actions must be carried out within the Stratonovich discretization scheme, but remarkably the action in the latter scheme does not pick up any additional contributions.  
 
However, it was explicitly shown \cite{MR05} that 
the direct application of the loop expansion for the dynamic action in Eq.~(\ref{eq2.6})
turns out to be incompatible with the fluctuation-dissipation relation (FDR) between 
the density correlation function $G_{\r \r} ({\bf r}, t)$ and the corresponding 
physical response function $R({\bf r}, t)$ (see (\ref{standardFDR}) below), which is the central element of the equilibrium dynamics.
Closely related to this, while in the additive Langevin equations
the noise-response $iG_{\r \hr}({\bf r}, t)$ is actually proportional 
to the physical response, 
that was shown to be not the case for the DK equation \cite{MR05} 
where the physical response is given by an unusual composite two-point function,
written down in Eq.~(\ref{standardFDR}) below.
ABL \cite{ABL06} further elucidated the origin of this inconsistency between the FDR and the RPT, and thereby provided an elegant way of resolving it
by focusing on the time-reversal (TR) symmetry of the dynamic action.
ABL then proposed the introduction of the conjugate pair of auxiliary fields 
$\{\th, \hth\}$ (defined below) to linearize the one of the TR transformations, 
which restores preservation of the FDR order by order in the loop expansion for the new form of the dynamic action incorporating $\{\th, \hth \}$. 
[KK08] developed a modified version of ABL's auxiliary field method. 

Yet another different scheme was recently proposed by two of the present authors in \cite{JvW11}, exploiting the reversibility of the dynamics to map the problem onto an effective quantum system.
This approach yields a reversible formulation, in which 
the case of noninteracting Brownian gas (NBG) ($U=0$ in (\ref{eq2.1}) 
is a Gaussian theory, but the correlation function in this approach is a four-point object, and thus new difficulties arise when 
attempting to perform a loop expansion. The formulation proposed in \cite{ABL06}, in [KK08], and in the present 
paper are more straightforward to analyze, since they are formulated in terms of two-body objects,
however, it will be shown that the case of NBG is a strongly non-Gaussian theory, giving rise to new difficulties.
More precisely, the new form of the action, when $U=0$, contains an infinite number of polynomial terms, 
which should be taken into account non-perturbatively. 
Instead, performing a loop-expansion amounts to 
developing in perturbation theory around a Gaussian ground state, whereas the NBG case is strongly non-Gaussian \cite{VCCK08} 
due to the multiplicative nature of the thermal noise acting on the density field.

\subsection{Time-reversal symmetries}

The dynamic action in Eq.~(\ref{eq2.6}) is invariant under the following two types of TR transformation for $\r $ and $\hr $.
 The U-transformation involves the free energy functional;
\beq
U:  \quad \left\{ \begin{array}{lll}
\r ({\bf r}, -t) & \to &  \r ({\bf r}, t) \ , \\
&&\\
\hr ({\bf r}, -t)  & \to & \displaystyle -\hr ({\bf r}, t)
+ \frac{i}{T} \frac{\d F[\r]}{\d \r ({\bf r},t)} \ .
\end{array} \right.
\label{eq2.7}
\eeq
Another transformation explicitly involves the time derivative;
\beq
T: \quad \left\{ \begin{array}{ll}
&\r({\bf r},-t) \to \r({\bf r},t) \ , \\
&\\
&\nabla \cdot \Big( \r({\bf r},-t) \nabla \hr({\bf r},-t) \Big) 
\to \displaystyle \\
& \quad \displaystyle \nabla \cdot \Big( \r({\bf r},t) \nabla \hr({\bf r},t)  \Big)
+\frac{i}{T}\partial_t \r({\bf r},t) \ .
\end{array} \right.
\label{eq2.8}
\eeq
The U-transformation was first written down by Janssen \cite{Ja79}, and 
the T-transformation appears first in \cite{ABL06}.
Important feature of these TR transformations is that 
both  are nonlinear: 
the former is nonlinear owing to the form of the entropic contribution $F_{\rm{id}}[\r]$, whereas the latter
due to the multiplicative nature  of the DK equation.
The standard FDR
follows from either of these TR transformations. 
It reads:
\beq \begin{split}
& \frac 1T \partial_{t} G_{\r\r}({\bf r}-{\bf r}',t-t') = R({\bf r}-{\bf r}',t'-t) - R({\bf r}-{\bf r}',t-t') \ , \\
& G_{\r\r}({\bf r}-{\bf r}';t-t') \equiv \la \d \r({\bf r},t) \d \r({\bf r}',t') \ra \ , \\
& R({\bf r}-{\bf r}',t-t') \equiv - \la \r({\bf r},t) \nabla^{'} \cdot \left[ \r({\bf r}',t')
 \nabla^{'} \hr({\bf r}',t') \right] \ra \ .
\end{split} 
\label{standardFDR}
\eeq
The problem is that the decomposed Gaussian and 
non-Gaussian components of the original dynamic action in Eq.~(\ref{eq2.6})
are separately invariant under neither transformations, which makes the direct loop expansion incompatible with the FDR.

Each TR transformation offers its own perturbation scheme preserving the FDR.
First of all, both transformations can be made linear by 
introducing extra sets of field variables.
In this regard, it is much more convenient to linearize the 
U-transformation since introducing a single set of extra fields 
(denoted by $\th$ and $\hth$) would suffice.
One then can apply the standard loop-expansion method for 
the new dynamic action incorporating the new fields $\th$ and 
$\hth$ (see (\ref{eq2.11}) below).

One can follow another  perturbation scheme, a potential
expansion method, which preserves the T-transformation.
In this scheme, one decomposes the original dynamic action into the interaction-free 
(corresponding to the NBG) and interaction 
contributions  instead of decomposing into the Gaussian and non-Gaussian parts;
\beq\begin{split}
{\cal S}[\r, \hr] & = S_{\rm{id}}[\r, \hr] + S_{\rm{int}} [\r, \hr] \ , \\
{\cal S}_{\rm{id}} [ \r, \hr ] & = \int_{{\bf r},t} \left\{  i\hr \left( \frac{}{} \partial_t \r - T\nabla^2  \r  \right) - T \r ( \nabla \hr)^2 \right\} \ , \\
{\cal S}_{\rm{int}} [ \r, \hr ] &= -\int_{{\bf r},t} i\hr ~ \nabla \cdot \left( \r \nabla \frac{\d F_{\rm{int}}[\r]}{\d \r} \right) \ .
\end{split}
\label{eq2.9} 
\eeq
Both $S_{\rm{id}} $ and $S_{\rm{int}} $ become separately invariant under the T-transformation.
Here although the interaction-free action $S_{\rm{id}}$ has the non-Gaussian cubic nonlinearity $-T\r (\nabla \hr)^2$, 
 the expansion of this nonlinearity in the bare Gaussian perturbation scheme 
is soon truncated due to the causality, namely the Gaussian average of the two hatted
 fields being zero. 
Thus the multiplicative thermal noise can be {\it exactly} treated in each order of potential expansion.
Here no extra set of variables is needed, and the NBG case ($U=0$) is trivially obtained.
But the calculation in the presence of interaction is much more involved than in the other 
scheme. The first steps of such an expansion are described in \cite{VCCK08}, but no 
one has yet completed the work utilizing the T-transformation.

\subsection{Linearization of the U-transformation}

In this work, we focus on the loop expansion method 
via the linearization of the U-transformation.
As stated earlier, it can be achieved by letting 
a new auxiliary  field $\th({\bf r},t)$ take up only
the nonlinear part of the U-transformation as 
\beq \begin{split}
\th ({\bf r},t) & = f(\d \r({\bf r},t) ) \equiv = \frac{1}{T} \frac{\d F_{\rm{id}}[\r]}{\d \r ({\bf r},t)}-\frac{\d \r ({\bf r},t)}{\r_0} \\
& = \ln \left( 1 + \frac{\d \r({\bf r},t)}{\r_0} \right) - \frac{\d \r({\bf r},t)}{\r_0}  \\
& = -\sum_{n=2}^{\io}\frac{(-1)^n}{n} \left(\frac{\d \r ({\bf r},t)}{\r_0}  \right)^n \ .
\end{split}
\label{eq2.10}
\eeq
Thus $\th$ field represents all the non-Gaussian density fluctuations coming from the entropic part of the free energy.

The resulting new dynamic action ${\cal S}[\r, \hr, \th, \hth ]$ is then decomposed 
into a Gaussian part ${\cal S}_g$ and a non-Gaussian part ${\cal S}_{\rm{ng}}$, defined as:
 \beq \begin{split}
& {\cal S} [  \r, \hr, \th, \hth ] = {\cal S}_{\rm{g}}[ \r, \hr, \th, \hth ] + {\cal S}_{\rm{ng}}[ \r, \hr, \th, \hth ] \ , \\
& {\cal S}_{\rm{g}} \equiv \int_{{\bf r},t} \left\{ \begin{array}{ll}
& \!\!\!\! \displaystyle i\hr \left[ \partial_t \r - T \nabla^2 \r -\r_0  T \nabla^2 \th - \r_0  \nabla^2 U \ast \d \r \right] \\
& \!\!\!\! \displaystyle - \r_0  T  ( \nabla \hr)^2 +  i \hth \th 
\end{array}\!\! \right\} \ , \\
& {\cal S}_{\rm{ng}} \equiv \int_{{\bf r},t} \left\{ \begin{array}{ll}
& \!\!\!\! \displaystyle i\hr \left[ - T \nabla \cdot \left( \d \r \nabla 
\left[ {\hat K} \ast \d \r \right] \right)
- T \nabla \cdot \left( \d \r \nabla \th \right) \right] \\
& \!\!\!\! \displaystyle - T \d \r ( \nabla \hr)^2 - i\hth f(\d\r) 
\end{array} \!\! \right\} \ , 
\end{split} 
\label{eq2.11}
\eeq
where $\ast$ denotes the space convolution, i.e., 
\beq
\left( {\hat K} \ast \d \r \right) ({\bf r},t) \equiv \int_{{\bf r}'} K(|{\bf r}-{\bf r}'|) \d \r({\bf r}',t) \ , 
\eeq
and $K$ is defined by:
\beq
K(|{\bf r}-{\bf r}'|) \equiv \frac{1}{\r_0} \d ({\bf r}-{\bf r}') + \frac{1}{T} U(|{\bf r}-{\bf r}'|) \ .
\label{eq2.13}
\eeq

In Eq.~(\ref{eq2.11}),  the field $\hth({\bf r},t)$ appears from the integral representation of the delta-functional constraint
$\d \big[ \th({\bf r},t)-f(\r({\bf r},t))\big]$ from Eq.~(\ref{eq2.10}), which 
 inevitably generates the nonpolynomial (logarithmic) nonlinearity $-i\hth f(\d \r) $, 
while the original action in Eq.~(\ref{eq2.6}) contains polynomial nonlinearities only.

Although Andreanov, Biroli, and Lef\`evre (ABL) \cite{ABL06} uses the same formal setting as the present work, the results they obtain have a rather different structure.
In particular, ABL uses a different definition of $\theta$ variable.
ABL's definition $\th_{ABL}=\delta F /\delta \r$ includes even the linear part of density fluctuation, whereas $\theta$ in Eq. (\ref{eq2.10}) explicitly separate the linear density fluctuation from $\delta F_{id} /\delta \r$.
Because of this difference, 
distinction between ABL and the present work persists through the forms of the new dynamic action, some of the FDRs, and the dynamic equations.
It is thus useful to keep in mind that 
 their (Fourier transformed) correlation functions 
$C_{\r\th}(k,t)$ and $C_{\th\th}(k,t)$ correspond respectively to 
$T\big(K(k)G_{\r\r}(k,t)+G_{\r\th}(k,t) \big)$ and $T \big( K(k)G_{\th\r}(k,t)+G_{\th\th}(k,t) \big)$ in our case. 
Even if this correspondence is taken into account, the resulting dynamic equations 
exhibit a subtle difference in both cases, particularly for the equations for 
$G_{\r\th} $ and $G_{\th\th}$.
At any rate, we tend to believe that ABL's analysis concerning the decay at long times of correlation functions of the theory is erroneous, and in particular, the equation 
for the nonergodicity parameter of $G_{\r\r}(k,t)$ (Eq. (97) of \cite{ABL06}), 
which is based on several hypotheses and on the neglect of the first memory integral
in the long-time limit (Eq. (88) of \cite{ABL06}), which would not be zero.
In this work we do not make such hypotheses, and our results are obtained without any assumptions other than the starting models. 
 
Now $ {\cal S}_{\rm{g}}$ and 
${\cal S}_{\rm{ng}}$ become  {\it separately} invariant under the {\it linearized} U-transformation 
\beq
U_L: ~
\left\{ \begin{array}{lll}
 \r ({\bf r}, -t)& = & \r ({\bf r}, t) \ ,  \\
 \hr ({\bf r}, -t)  &=&  -\hr ({\bf r}, t) +
 i \big( {\hat K} \ast \d \r \big)({\bf r},t)+i\th({\bf r},t) \ , \\
\th ({\bf r}, -t) & = & \th ({\bf r}, t) \ ,  \\
\hat \th ({\bf r}, -t)  &=&  \hat \th ({\bf r}, t)
+i \partial_t \r ({\bf r},t) \ ,
\end{array} \right.
\label{eq2.14}
\eeq
Here the modulus of the associated transformation matrix $O$ is unity ($\det O=-1$). 
That the transformation of $\hr$ field involves 
not only $\r$ but also $\th$ fields reflects non-Gaussian contributions of the entropic  free energy. The response field $\hth$  acts as the time derivative of density fluctuation for time-reversal invariance since  it couples with a function of density fluctuation. 

Now due to the separate invariance of the new dynamic action 
under the $U_L$ transformation,
the RPT for the new dynamic action  is guaranteed to preserve the FDR order by order.

However, it is obvious from the form of the action (\ref{eq2.11}) that the 
NBG case is not a Gaussian theory, since setting 
$U=0$ in the action does not cancel its non-Gaussian part. An order by order loop expansion
is thus not guaranteed to treat correctly the NBG.

Until now, we have recalled the preliminary steps of the work presented in [KK08], for the sake of clarity, but in the following
some results taken from this previous work will be recalled without proof, for the sake of compactness.

\section{Dynamic equations}
\label{dynamic-equ}

The linear transformation under time reversal in Eq.~(\ref{eq2.14})
dictates the following linear FDRs between the correlation and response functions:
\beq
\left\{ \begin{array}{ll}
G_{\r\hr}(k,t) & =  i \Th(t) \left( K(k) G_{\r\r}(k,t) + G_{\r\th}(k,t) \right) \ , \\
G_{\r \hth}(k,t) & =  i \Th(t) \partial_t G_{\r \r}(k,t) \ , \\
 G_{\th\hr}(k,t) & =  i \Th(t) \left( K(k) G_{\th\r}(k,t) + G_{\th\th}(k,t) \right) \ ,  \\
G_{\th\hth}(k,t) & = i \Th(t) \partial_t G_{\th\r}(k,t) + i \d(t)  \ ,
\end{array} \right.
\label{eq3.1}
\eeq
where $K(k) \equiv 1/\r_0 + U(k)/T$ is the Fourier transform of 
$K(r)$ defined in Eq.~(\ref{eq2.13}).
Due to the nature of $\th$ field, $G_{\r\th}(k,t)$ in the first FDR is a non-Gaussian contribution (from the entropic part of the free energy) to the density correlation. 
Since $\hth$ field acts as time derivative of density fluctuation, 
the second FDR in (\ref{eq3.1}) is nothing but the standard FDR, the first line of (\ref{standardFDR}) with the physical response now reduced to 
$R({\bf r},t)=iG_{\r\hth}({\bf r},t)/T$.
The additional Dirac delta term in $G_{\th\hth}$ in (\ref{eq3.1}) does not come from the FDR
but is a contribution coming directly from the bare correlator given by the action 
in Eq.~(\ref{eq2.11}),
which is not renormalized by any loop diagrams as can be seen by direct inspection.
This term reflects the fact that the Lagrange multiplier $\hth$ acts also 
at equal times, and thus affects the equilibrium correlations in addition to dynamic ones.
[KK08] missed out this last term, which significantly affects the analysis given in this previous work.
In particular, the overlooked delta function generates additional terms in the equations of motion 
for the correlation and response functions, and thereby stymies the nonperturbative analysis given 
in Section 3.9 of [KK08]. The new analysis given below remedies these erroneous results.
We note that such a term was correctly taken into account in the previous work of 
ABL \cite{ABL06}.

The self-energies are defined as the functional inverse of the propagators 
via the Schwinger-Dyson (SD) equation:
\beq
\mathbb{1} ~ \d(t) = \left( \mathbf{G_0}^{-1} \otimes \mathbf{G} \right) (k,t) - \left( \mathbf{\S} \otimes \mathbf{G} \right)(k,t) \ ,
\label{eq3.3}
\eeq
where $\mathbf{G_0}^{-1}$ is the inverse of the bare propagator, which is simply read off from the dynamical action, $\mathbf{G}$ and 
$\mathbf{\S}$ are $4 \times 4$ matrices containing all correlators and self-energies respectively, and 
$\mathbb{1}$ is a $4 \times 4$ identity matrix.
The symbol $\otimes$ stands for time convolution.
The FDRs holds also for the self-energies, and read:
\beq
\left\{ \begin{array}{ll}
\S_{\hr\r}(k,t) & = i \Th(t) \left( - K(k) \S_{\hr\hr}(k,t) + \partial_t \S_{\hr\hth}(k,t) \right) \ , \\
\S_{\hr\th}(k,t) & = - i \Th(t) \S_{\hr\hr}(k,t) \ , \\
\S_{\hth\r}(k,t) & = i \Th(t) \left( - K(k) \S_{\hth\hr}(k,t) + \partial_t \S_{\hth\hth}(k,t) \right) \ , \\
\S_{\hth\th}(k,t) & = - i \Th(t) \S_{\hth\hr}(k,t) \ , \\
\S_{\hth\hr}(k,t) & = - \S_{\hth\hr}(k,-t).
\end{array}\right.
\label{eq3.2}
\eeq
The FDRs in Eqs.~(\ref{eq3.1}) and (\ref{eq3.2}) are employed to derive the dynamic equations 
 for the correlation functions given in the next section.

\subsection{Equations for the correlation functions}

The dynamic equations for the correlation and response functions are formally given by
the matrix SD equation in Eq.(\ref{eq3.3}).

We first write down the dynamic equations for the correlation functions:
\beq \begin{split} 
& \partial_t G_{\r\r}(k,t) = - \r_0 T k^2 \left[ \frac{}{} K(k) G_{\r\r}(k,t) + G_{\th\r}(k,t) \right] \\
& + \int_0^t \! ds \left( \begin{array}{ll}
& \displaystyle \!\!\!\! \S_{\hr\hr}(k,t-s) \left[ \frac{}{} \! K(k) G_{\r\r}(k,s) + G_{\th\r}(k,s) \right] \\
& \\
& \displaystyle \!\!\!\! - \S_{\hr\hth}(k,t-s)\partial_s G_{\r\r}(k,s) 
\end{array} \!\! \right) \ , 
\end{split}
\label{eq3.4}
\eeq

\beq\begin{split} 
& G_{\th\r}(k,t) = \S_{\hth\hth}(k,0) G_{\r\r}(k,t) \\
& + \int_0^t \! ds \left( \begin{array}{ll}
& \displaystyle \!\!\!\! \S_{\hth\hr}(k, t-s) \left[ \frac{}{}\! K(k) G_{\r\r}(k,s) + G_{\th\r}(k,s) \right] \\
& \\
& \displaystyle \!\!\!\! - \S_{\hth\hth}(k, t-s) \partial_s G_{\r\r}(k,s)
\end{array} \!\! \right) \ , 
\end{split}
\label{eq3.5}
\eeq

\beq\begin{split} 
& \partial_t G_{\r\th}(k,t) = - \r_0 T k^2 \left[ \frac{}{}\! K(k)G_{\r\th}(k,t) + G_{\th\th}(k,t) \right] \\
& \phantom{\partial_t G_{\r\th}(k,t) = } - \S_{\hr\hth}(k,t) \left[ \frac{}{} \! K(k) G_{\r\r}(k,0) + G_{\th\r}(k,0) \right] \\
& + \int_0^t \! ds \left( \begin{array}{ll}
& \displaystyle \!\!\!\! \S_{\hr \hr}(k,t-s) \left[ \frac{}{} K(k) G_{\r\th}(k,s) + G_{\th\th}(k,s) \right] \\
& \\
& \displaystyle \!\!\!\! - \S_{\hr\hth}(k,t-s) \partial_s G_{\r\th}(k,s) 
\end{array} \!\! \right) \ , 
\end{split}
\label{eq3.6}
\eeq

\beq \begin{split} 
& G_{\th\th}(k,t) = \S_{\hth\hth}(k,0) G_{\r\th}(k,t) \\
& \phantom{G_{\th\th}(k,t) =} - \S_{\hth\hth}(k,t) \left[ \frac{}{}\! K(k) G_{\r\r}(k,0) + G_{\th\r}(k,0) \right] \\
& + \int_0^t \! ds \left( \begin{array}{ll}
& \displaystyle \!\!\!\! \S_{\hth\hr}(k,t-s) \left[ \frac{}{}\! K(k) G_{\r\th}(k,s) + G_{\th\th}(k,s) \right] \\
& \\
& \displaystyle \!\!\!\! - \S_{\hth\hth}(k, t-s) \partial_s G_{\r\th}(k,s) 
\end{array} \!\! \right) \ .
\end{split}
\label{eq3.7}
\eeq
The second terms on the right hand sides of Eqs.~(\ref{eq3.6}) and (\ref{eq3.7}) are the ones generated by the presence of the delta function in the last member of Eq.~(\ref{eq3.1}).
These new terms turn out to invalidate the nonperturbative 
results given in Sec. 3.9 of [KK08]. 
As a result, one cannot obtain a single closed equation for $G_{\r\r}(k,t)$ alone.
Instead, one should face a set of coupled equations for the correlation
functions involving $\th$ variables. 
The very same situation is observed in the formulation
of Jacquin and van Wijland \cite{JvW11}, and it is a difficulty that seem to naturally occur when pairs of extra fields are introduced.

\subsection{Static input}

Since we are dealing with the dynamics of fluctuations around the 
equilibrium state, 
the static information prescribed by the free energy $F[\r]$ in Eq.~(\ref{eq2.3}) 
should be consistent with the initial conditions for the dynamic equations of motion for the correlation functions. 
Injecting the causal forms of the correlators in the SD equation~(\ref{eq3.3}), 
time derivatives contained in the inverse propagator will act on
the Heaviside functions and produce Dirac delta functions. Now assuming that only $G_{\th\hth}$ contains a delta function, 
and that no self-energy gets such a term, we can readily equate the delta functions 
in Eq.~(\ref{eq3.3}) 
 we obtain directly:
\beq \begin{split} 
& \begin{pmatrix}
{\cal X}(k,0) & \quad {\cal Y}(k,0) & 0 & 0 \\
0 & 1 & 0 & 0 \\
0 & 0 & \quad {\cal X}(k,0) & \quad \dot{G}_{\r\r}(k,0) + \r_0 T k^2 \\
0 & 0 & 0 & 1
\end{pmatrix} \\
& = \begin{pmatrix}
1 & 0 & 0 & 0 \\
0 & 1 & 0 & 0 \\
0 & 0 & 1 & 0 \\
0 & 0 & 0 & 1
\end{pmatrix} \ , 
\end{split}
\label{eq3.8}
\eeq
where the dot stands for time derivative.
In Eq.~(\ref{eq3.8}), we have defined:
\beq
\left\{ \begin{array}{ll}
& {\cal X}(k,t) \equiv K(k) G_{\r\r}(k,t) + G_{\r\th}(k,t) \ , \\
& \\
& {\cal Y}(k,t) \equiv K(k) G_{\r\th}(k,t) + G_{\th\th}(k,t) \ . 
\end{array} \right.
\label{eq3.9}
\eeq
We thus obtain  three constraints, which will be shown to be fully consistent with the rest of the dynamical equations:
\beq \begin{split}
& {\cal X}(k,0) = 1 \ , \\
& {\cal Y}(k,0) = 0 \ , \\
& \dot{G}_{\r\r}(k,0)= - \r_0 T k^2 \ .
\end{split}
\label{eq3.10}
\eeq

Setting $t=0^+$ in Eqs.~(\ref{eq3.4})-(\ref{eq3.7}) gives
\beq \begin{split}
\dot G_{\r\r}(k,0) & = -\r_0 T k^2 {\cal X}(k,0) \ , \\
G_{\th\r}(k,0) &=  \S_{\hat\theta \hat\theta}(k,0) 
G_{\rho\rho}(k,0) \ , \\
\dot G_{\rho\theta}(k,0)&= - \r_0 T k^2 {\cal Y}(k,0) \ , \\
G_{\theta\theta}(k,0)&= - \S_{\hat\theta \hat\theta}(k,0) 
 K(k) G_{\r \r}(k,0) \ .
 \end{split} 
\label{eq3.11}
\eeq

Since the system is assumed to be in equilibrium, we 
feed the equilibrium correlation, 
which is the static structure factor of the liquid:
\beq
G_{\rho \rho}(k,0) = \r_0 S(k)
\label{eq3.12}
\eeq
Injecting this and Eq.~(\ref{eq3.10}) directly into Eq.~(\ref{eq3.11}), 
we obtain the important, non-perturbative relation:
\beq \begin{split}
\S_{\hth\hth}(k,0) & = \left( \frac{1}{\r_0 S(k)} - K(k) \right) = - \left( c(k) + \frac{U(k)}{T} \right) \ , \\
G_{\r \th}(k, 0) & = 1 - \r_0 S(k) K(k) \ , \\
G_{\th\th}(k,0) & = - K(k) \left( \frac{}{}\! 1 - \r_0 S(k) K(k) \right) \ ,
\end{split}
\label{eq3.13}
\eeq
where we expressed in the first line the static structure factor $S(k)$ in terms of 
the direct correlation function \cite{hansen} $c(k)$ using 
\beq
\frac 1{ S(k)} = 1 - \r_0 c(k) \ .
\nonumber
\eeq

\subsection{Analysis of Dynamic equations}

In [KK08], overlooking the extra delta function contribution
to $G_{\th\hth}(k,t)$ in Eq.~(\ref{eq3.1}), 
the authors {\it falsely} obtained  the relation 
$$
\frac{G_{\r\th}(k,t)}{G_{\r\th}(k,0)} = \frac{G_{\r\r}(k,t)}{\r_0 S(k)} \ .
$$
This incorrect relation turns  the first FDR in Eq.~(\ref{eq3.1}) into
the (false) linear FDR between $G_{\r\hr}(k,t) $ and $G_{\r\r}(k,t) $
$$
G_{\r\hr}(k,t) = i \Th(t) \frac{1}{\r_0 S(k)} G_{\r\r}(k,t) \ .
$$
This erroneous FDR led to a closed equation for $G_{\r\r}(k,t) $ alone. 
This is not so. One really has to face a coupled  set of dynamic 
equations for the correlation functions $G_{\r\r}(k,t)$, $G_{\r\th}(k,t)$ (or $G_{\th\r}(k,t) $), 
and $G_{\th\th}(k,t)$.
We write them down again incorporating the static input;
\beq \begin{split}
& \partial_t G_{\r\r}(k,t) = - \r_0 T k^2 {\cal X}(k,t) \\
& + \!\! \int_0^t \!\!\! ds \! \left[ \frac{}{}\! \S_{\hat\rho \hat\rho}(k,t-s) {\cal X}(k,s) - \S_{\hr \hth}(k,t-s)\partial_s G_{\r\r}(k,s) \right],
\end{split}
\label{eq6.1}
\eeq

\beq \begin{split}
& {\cal X} (k,t) = \frac{1}{\r_0 S(k)}G_{\r\r}(k,t) \\
& + \!\! \int_0^t  \!\!\! ds \! \left[ \frac{}{}\! \S_{\hth \hr}(k,t-s) {\cal X}(k,s) - \S_{\hth \hth}(k,t-s)\partial_s G_{\r\r}(k,s) \right] ,
\end{split}
\label{eq6.2}
\eeq

\beq \begin{split}
& \partial_t G_{\r\th}(k,t) = - \r_0 T k^2 {\cal Y}(k,t)  - \S_{\hr\hth}(k,t) \\
& + \!\! \int_0^t \!\!\! ds \! \left[ \frac{}{}\! \S_{\hr\hr}(k,t-s) {\cal Y}(k,s) - \S_{\hr\hth}(k,t-s) \partial_s G_{\r\th}(k,s) \right] , 
\end{split}
\label{eq6.3}
\eeq

\beq \begin{split}
& {\cal Y}(k,t) = \frac{1}{\r_0 S(k)} G_{\r\th}(k,t) - \S_{\hth\hth} (k,t) \\
& + \!\! \int_0^t \!\!\! ds \! \left[ \frac{}{}\! \S_{\hth \hr}(k, t-s){\cal Y}(k,s) - \S_{\hth\hth}(k, t-s) \partial_s G_{\r\th}(k,s) \right] .
\end{split}
\label{eq6.4}
\eeq
The above equations~(\ref{eq6.1})-(\ref{eq6.4}) form a coupled set of 
FDR-preserving dynamic equations for 
the correlation functions $G_{\r\r}(k,t)$, 
$G_{\r\th}(k,t)$ (or $G_{\th\r}(k,t) $), and $G_{\th\th}(k,t) $ since
the response functions appearing in the self-energies  can be replaced by the corresponding correlation functions via 
the linear FDRs in Eq.~(\ref{eq3.1}). 
The second terms in the right hand sides of Eqs.~(\ref{eq6.3}) and (\ref{eq6.4}) 
are those generated by the overlooked delta function contribution
in the last member of Eq.~(\ref{eq3.1}).
Due to these terms, 
the relation $G_{\r\hr}=i \Th(t) G_{\r\r}(k,t)/\r_0 S(k)$ is {\it invalid}
 and consequently one cannot obtain the single closed equation for 
$G_{\r\r}(k,t)$.
These equations with the explicit expressions for the self-energies, 
could be solved numerically with the given initial conditions. 

Let us rearrange the dynamic equations~(\ref{eq6.1})-(\ref{eq6.4}) for analysis.
Subtracting Eq.~(\ref{eq6.2}) multiplied by $\r_0 T k^2$ from Eq.~(\ref{eq6.1}), we obtain 
\begin{align}
& \partial_t G_{\r\r}(k,t) = - \frac{T k^2}{S(k)} G_{\r\r}(k,t) \label{eq6.5} \\
& + \int_0^t  ds \left[ \frac{}{}\! \S_1(k,t-s) {\cal X}(k,s) - \S_2(k, t-s) \partial_s G_{\r\r}(k,s) \right] \nonumber ,
\end{align}
where $\S_1(k,t)$ and $\S_2(k,t)$ are defined as
\beq
\S_1(k,t) \equiv \S_{\hr\hr}(k,t)
- \r_0 T k^2 \S_{\hth\hr}(k,t) \ , 
\label{defSigma1}
\eeq
\beq
\S_2(k,t) \equiv \S_{\hr\hth}(k,t)
- \r_0 T k^2 \S_{\hth\hth}(k, t) \ .
\label{defSigma2}
\eeq

Note that the first memory integral in Eq.~(\ref{eq6.5}) is folded onto ${\cal X}(k,t)$ 
instead of the time derivative of $G_{\r\r}(k,t)$ as in the second one. 
One recognizes from the form of Eq.~(\ref{eq6.1}) that one can make the first integral 
folded into the time derivative of $G_{\r\r}(k,t)$ with the new memory function ${\cal M}(k, t)$;
\beq
\int_0^t \!\!\! ds ~ \S_1(k, t-s) {\cal X} (k,s) \equiv - \!\! \int_0^t \!\!\! ds ~ {\cal M}(k, t-s) \, \partial_s G_{\r\r}(k,s)
\label{eq6.8}
\eeq
where the memory kernel ${\cal M}(k,t) $ obeys the following exact equation (see Appendix 
\ref{app_derivation-kernels}):
\beq \begin{split}
\r_0 T k^2 {\cal M} (k,t) = & \S_1(k, t) - \int_0^t  ds ~ \S_1 (k,t-s) \S_{\hth\hr} (k, s) \\
& + \int_0^t  ds ~ \S_{\hr\hr} (k,t-s) {\cal M} (k, s) \ .
\end{split}
\label{eq6.9}
\eeq
This rearrangement is superficially analogous to the one employed in the irreducible memory 
function formulation \cite{CH87,Ka95} in the projection operator approach. In our case, it is not
an operation that simplifies the diagrammatic structure of the memory kernel, but
it is rather a projection onto the subspace of density, factoring out the $\theta$ correlations, 
which is a necessary step to make contact with the SMCT.
We note that the folding onto the time derivative of $G_{\r\r}(k, t)$ causes a fundamental change in
the structure of the loop expansion by inevitably bringing higher order contributions to the new memory
function ${\cal M}(k,t) $. That is,  in addition to the one-loop term, $\S_1(k, t)/\r_0 T k^2$, 
higher order terms are generated by iterating Eq.~(\ref{eq6.9}) for ${\cal M}(k,t)$.
As shown below, this lowest order term $\S_1(k, t)/\r_0 T k^2$ gives SMCT, but the full one loop calculation contains 
these higher order contributions beyond SMCT (which gives rise to ENE transition) such that this 'extra' part of one loop gives rise to a smearing out of SMCT.

Substituting Eq.~(\ref{eq6.8}) into Eq.~(\ref{eq6.5}), one can rewrite Eq.~(\ref{eq6.5}) as
the following exact equation
\begin{align}
\partial_t G_{\rho\rho}(k,t) = & - \frac{T k^2}{S(k)} G_{\rho \rho}(k,t)  
\label{eq6.10} \\
& - \int_0^t  ds \, \big( {\cal M} +\S_2 \big)(k,t-s) \partial_s G_{\rho \rho}(k,s) \nonumber \ .
\end{align}
We must stress that Eq.~(\ref{eq6.10}) is {\it not} a closed equation for $G_{\r\r}(k,t)$ alone  since the memory kernels ${\cal M}(k,t)$ and $\S_2(k,t)$ 
will involve both $G_{\th\r}(k,t)$ and $G_{\th\th}(k,t)$ as well.

In the same manner, one can also rearrange Eq.~(\ref{eq6.2}) as
\begin{align}
{\cal X}(k,t) = & \frac{1}{\r_0 S(k)} G_{\r\r}(k,t) \label{eq6.11} \\
& - \int_0^t  ds\,  \Big( {\cal N} + \S_{\hat\th \hat\theta}\Big) (k,t-s)  \partial_s G_{\rho \rho}(k,s) \nonumber \ .
\end{align}
Here the new memory kernel ${\cal N}(k,t)$, being defined as
\beq
\int_0^t \!\!\! ds ~ \S_{\hth \hr}(k, t-s) {\cal X} (k,s) \equiv - \!\! \int_0^t \!\!\! ds ~ {\cal N}(k, t-s) \partial_s G_{\rho\rho}(k,s) ,
\label{defkernelN}
\eeq
obeys (see Appendix \ref{app_derivation-kernels})
\begin{align}
& \r_0 T k^2 {\cal N} (k,t) = \S_{\hth\hr}(k, t)  \label{equationforkernelN} \\ 
& - \int_0^t  ds \left( \S_{\hth \hr} (k,t-s) \S_{\hth \hr} (k, s) - \S_{\hat\rho \hat\rho} (k,t-s) {\cal N} (k, s) \right). \nonumber 
\end{align}

Eqs.~(\ref{eq6.3}) and (\ref{eq6.4}) can be rewritten as:
\begin{align}
& \partial_t G_{\rho\th}(k,t) = - \frac{T k^2}{S(k)} G_{\rho \th}(k,t) -\S_2(k,t) \label{eq6.14} \\
& + \int_0^t \!\! ds \left[ \frac{}{} \! \S_1(k,t-s) {\cal Y} (k,s) - \S_2(k, t-s) \partial_s G_{\rho \th}(k,s) \right] \ , \nonumber
\end{align}
\begin{align}
& {\cal Y}(k,t) = \frac{1}{\r_0 S(k)} G_{\r\th}(k,t) -\S_{\hth\hth}(k,t) \label{eq6.15} \\
& + \!\! \int_0^t  \!\!\! ds \left[ \frac{}{}\! \S_{\hth\hr}(k,t-s) {\cal Y} (k,s) -  \S_{\hth \hth}(k,t-s) \partial_s G_{\r\th}(k,s) \right] \ . \nonumber
\end{align}

\section{One-loop approximation and SMCT}

In order to obtain an approximation for the complete dynamical problem,
we calculate the one-loop expressions for the self-energies.
We simply recall the results already obtained in [KK08], without proof.
The simplest self-energy $\S_{\hth\hth}(k,t)$ is given by
\beq
\S_{\hth \hth} (k,t) = -\frac{1}{2\r_0^4} \int_{\bf q} G_{\r\r}(q,t) G_{\r\r}(|{\bf k}-{\bf q}|,t) \ ,
\label{eq6.16}
\eeq
where $\int_{\bf q} \equiv \int d^3 {\bf q}/(2\pi)^3 $.
$\S_2$ is given at one-loop by:
\begin{align}
& \S_2 (k, t) =  \label{eq6.19} \\
& - \frac{T}{\rho^2_0} \int_{\bf q} \! {\bf k}\cdot {\bf q} \! \left[ \frac{U(q)}{T} G_{\rho\rho}(q,t) + G_{\th\rho}(q,t) \right] G_{\rho\rho}(|{\bf k}-{\bf q}|,t) . 
\nonumber
\end{align}
Finally the one-loop expression of $\S_1(k,t)$ reads:
\beq \begin{split}
& \S_1(k, t) = \frac{T^2k^2}{\r_0} \int_{\bf q} \, {\bf k}\cdot {\bf q} \, G_{\th \rho}(q,t) G_{\rho\rho}(|{\bf k}-{\bf q}|,t) \\
& + T^2 \int_{\bf q} \, {\bf k}\cdot {\bf q} \, \, {\bf k}\cdot \big( {\bf k}-{\bf q} \big) G_{\th \rho}(q,t) G_{\rho\th}(|{\bf k}-{\bf q}|,t) \\
& + T^2 \int_{\bf q} \, \big({\bf k}\cdot {\bf q} \big)^2  G_{\th \th}(q,t) G_{\rho\r}(|{\bf k}-{\bf q}|,t) \\
& + \frac{1}{2} \int_{\bf q} \,  \Big[ {\bf k}\cdot {\bf q}\, U(q)+ {\bf k}\cdot \big( {\bf k}-{\bf q} \big)\, U(|{\bf k}-{\bf q}|) \Big]^2 
G_{\rho\rho} G_{\rho\rho} \\
& + \frac{Tk^2}{2\r_0} \int_{\bf q} \, 
 \Big[ {\bf k}\cdot {\bf q}\, U(q)
+ {\bf k}\cdot \big( {\bf k}-{\bf q} \big)\, U(|{\bf k}-{\bf q}|) \Big]
G_{\rho\rho} G_{\rho\rho} \\
& + 2T \int_{\bf q} \, {\bf k}\cdot {\bf q} \,
 \Big[ {\bf k}\cdot {\bf q}\, U(q)
+ {\bf k}\cdot \big( {\bf k}-{\bf q} \big)\, U(|{\bf k}-{\bf q}|) \Big] \\
& \phantom{2T \int_{\bf q} \, \qquad} G_{\th \rho}(q,t) G_{\rho\rho}(|{\bf k}-{\bf q}|,t) \ .
 \end{split}
 \label{eq6.22}
\eeq

Thus, the FDR-preserving one-loop theory is represented by a coupled set of dynamic equations: 
Eqs.~(\ref{eq6.10}) with Eq.~(\ref{eq6.9}), Eq.~(\ref{eq6.11}) with Eq.~(\ref{equationforkernelN}), and Eqs.~(\ref{eq6.14})-(\ref{eq6.15}), 
with the one-loop self-energies given by Eqs.~(\ref{eq6.16}), (\ref{eq6.19}), and (\ref{eq6.22}).
In the following we show how the SMCT equation can be retrieved within the one-loop theory.
We emphasize, however, that the one-loop theory is {\it not} tantamount to the SMCT; we wil show that 
the full one-loop theory does not support the dynamic transition into 
nonergodic phase predicted by the SMCT, i.e., the equilibrium dynamics 
remains always {\it ergodic}  within the full one-loop theory.

\subsection{Contact with the SMCT}

We have now completed the description of the dynamical theory, and shown that a simple one-loop approximation is not 
sufficient to obtain the SMCT equation. In order to do so, one must force the dynamical equations to be expressed
in function only of the density-density correlator $G_{\r\r}(k,t)$.

We thus introduce a book-keeping parameter $\l$ in front of each self-energy i.e. we make the replacement:
\beq
S_{\rm{ng}} \to \sqrt{\l} S_{\rm{ng}} \ , 
\eeq
and use $\l$ as an organizing device for the diagrammatic expansion. We will expand the equations to lowest order in 
$\l$, and evaluate the result at $\l=1$ at the end. 
This is of course not an expansion in powers of a small
parameter, but neither is the loop expansion!
This procedure of introducing a fictitious expansion parameter in the calculation, then equating it to one originates from
quantum field theory, where an expansion in powers of $\hbar$ produces a loop expansion. In statistical field theory, such a 
small parameter is usually absent (with the exception of $\mathcal{O}(N)$ models for example, where expansions in powers of $1/N$ can be performed), 
and the loop expansion is produced
by placing a $1/\l$ factor in front of the action, then organizing the diagrammatic expansion in powers of $\l$, before setting $\l$ to 1 at the end of the calculation.

From their expressions in Eqs.~(\ref{defSigma1}) and (\ref{defSigma2}) we are thus bound to find:
\beq \begin{split}
& \S_1 = \l ~ \S_1^{(1)} + \l^2 ~\S_1^{(2)} + \mathcal{O}( \l^3 ) \ , \\
& \S_2 = \l ~ \S_2^{(1)} + \l^2 ~ \S_2^{(2)} + \mathcal{O}( \l^3) \ ,
\nonumber
\end{split} 
\eeq
and thus we find from Eq.~(\ref{eq6.9}) 
\beq
\mathcal{M}(k,t) = \frac 1{\r_0 T k^2} \S_1^{(1)}(k,t) ~ \l + \mathcal{O}(\l^2) \ .
\eeq
Thus to order one in $\l$, we obtain from (\ref{eq6.10}) the following equation:
\begin{align}
& \partial_t G_{\r\r}(k,t) = -\frac{T k^2}{S(k)} G_{\r\r}(k,t) \label{eq6.33} \\
& - \l \int_0^t  ds \, \left( \frac{\S_1^{(1)}}{\r_0 T k^2}  +  \S_2^{(1)} \right)(k, t-s)  \, \partial_s G_{\r\r}(k,s) + \mathcal{O}(\l^2) . \nonumber
\end{align}

This equation is still not a closed equation 
for $G_{\r\r}(k,t)$ since the self-energies $\S_1^{(1)}$ and $\S_2^{(1)}$ depend on 
$G_{\th\r}(k,t)$ and $G_{\th\th}(k,t)$ as well.
But with the same reasonings, Eqs.~(\ref{eq6.11}--\ref{eq6.15}) show that:
\beq \begin{split}
& G_{\r \th}(k,t) = \left(\frac 1{\r_0 S(k)} - K(k) \right) G_{\r\r}(k,t) + \mathcal{O}(\l) \ , \\
& G_{\th\th}(k,t) = \left(\frac 1{\r_0 S(k)} - K(k) \right)^2 G_{\r\r}(k,t) + \mathcal{O}(\l) \ . 
\nonumber
\end{split} 
\eeq
Thus at order one in $\l$, the self-energies
involve only the density correlation function $G_{\r\r}(k,t)$, and we recover the false FDR obtained in [KK08].
Within this level of approximation, all results in [KK08] thus hold.

Eq.~(\ref{eq6.19}) shows that:
\beq \begin{split}
& \S_2^{(1)}(k,t)= \label{eq6.34} \\
& \frac{T}{2 \rho^2_0} \int_{\bf q} \Big[ {\bf k}\cdot {\bf q}\, c(q)
+ {\bf k}\cdot \big( {\bf k}-{\bf q} \big)\, c(|{\bf k}-{\bf q}|) \Big]  G_{\rho\rho} G_{\rho\rho} \ 
\end{split}
\eeq
Note that in Eq.~(\ref{eq6.34}) the interaction potential $U$ cancels out, 
and the direct correlation function of the liquid, $c(q)$, naturally emerges through the static structure factor $1/S(q)=1-\r_0 c(q)$.

Likewise, Eq.~(\ref{eq6.22}) reduces to 
\beq \begin{split}
& \frac{1}{\r_0 T k^2} \S_1^{(1)}(k,t) = \label{eq6.35} \\
& \frac{T}{2\r_0 k^2} \int_{\bf q} \,  \Big[ {\bf k}\cdot {\bf q}\, c(q) 
+ {\bf k}\cdot \big( {\bf k}-{\bf q} \big)\, c(|{\bf k}-{\bf q}|) \Big]^2 G_{\r\r} G_{\r\r} \\
& - \frac{T}{2\r_0^2} \int_{\bf q} \,  \Big[ {\bf k}\cdot {\bf q}\, c(q)
+ {\bf k}\cdot \big( {\bf k}-{\bf q} \big)\, c(|{\bf k}-{\bf q}|) \Big] G_{\r\r} G_{\r\r} \ . 
\end{split}\eeq
With Eqs.(\ref{eq6.34}) and (\ref{eq6.35}), Eq.~(\ref{eq6.33}) yields the SMCT equation 
when evaluated to first order in $\l$ then at $\l=1$:
\begin{align}
& \partial_t G_{\rho\rho}(k,t) = - \frac{T k^2}{S(k)} G_{\rho \rho}(k,t) \nonumber \\
& \hspace{2cm} - \int_0^t  ds \, M_{\rm{MCT}}(k, t-s) \partial_s G_{\rho \rho}(k,s) \ , \nonumber \\
& M_{\rm{MCT}}(k,t) = \label{eq6.36} \\
& \frac{T}{2\r_0 k^2} \int_{\bf q} \,  \Big[ {\bf k}\cdot {\bf q}\, c(q)
+ {\bf k}\cdot \big( {\bf k}-{\bf q} \big)\, c(|{\bf k}-{\bf q}|) \Big]^2 G_{\r\r} G_{\r\r} \ . \nonumber
\end{align}
It is remarkable  that  the inter-particle potential $U(q)$, appearing in the original DK equation, completely cancels out, and 
in the final result the direct correlation function $c(q)$ emerges, which is  due to the consistency between the dynamics and the statics.

In addition, we point out that if from the outset one takes a coarse-graining  point of view
and uses for  the form of the free energy functional 
the Ramakrishnan-Yousouff form, i.e., $U(r)=-Tc(r)$ in Eq.~(\ref{eq2.3}), 
we would obtain the SMCT result in much simpler way.
With $U(k)=-Tc(k)$, we have from the static input that
$\S_{\hth\hth}(k,0)=-(c(k)+U(k)/T)=0$, and hence
$G_{\th\r}(k,0)=0 $ and $G_{\th\th}(k,0)=0 $.
Thus, in order to obtain the SMCT result, when we 
plug $U(q)=-Tc(q)$ and $G_{\th\r}(q,t)\approx 0$ into
(\ref{eq6.19}), we immediately obtain Eq.~(\ref{eq6.34}).
In the same manner, when we substitute
$U(q)=-Tc(q)$, $G_{\th\r}(q,t) \approx 0 $ and 
$G_{\th\th}(q,t) \approx 0$ into Eq.~(\ref{eq6.22}), we obtain Eq.~(\ref{eq6.35}). 
We would thus have the SMCT result much more easily.

The straight one-loop expansion arising from the FDR-preserving one-loop expansion requires to be further worked out to yield 
the SMCT equation: coming back to the full dynamical equations~(\ref{eq6.9}--\ref{eq6.15}), we can see that the SMCT 
result is retrieved  when 
\begin{itemize}
\item  the memory (convolution) contributions to ${\cal M}(k,t)$ are ignored, and 
\item the memory contributions to ${\cal X}(k,t)$ and ${\cal Y}(k,t)$ are also ignored.
\end{itemize}
Both operations can easily be seen to be equivalent to the expansion in powers of $\l$ that we performed.
Taking into account the full set of the dynamic equations with the one-loop self-energies  is actually  found to {\it smear out} 
the sharp transition predicted by the SMCT, making the equilibrium dynamics 
always ergodic, which is shown in the next subsection.

\subsection{Absence of ENE transition}

Here we examine the long-time-limit behavior of the dynamic equations
to show that the full dynamic equations with the one-loop
self-energies do not support dynamic transition into
the nonergodic phase predicted by the SMCT.

We first introduce the notations for the long-time limit of 
the various quantities 
\beq \begin{split}
&f_{\r\r}(k) \equiv  \lim_{t \to \infty} G_{\r\r}(k,t), \\
& \sigma_{\hr\hr}(k) \equiv \lim_{t \to \infty} \Sigma_{\hr\hr}(k,t), \\
& x(k) \equiv \lim_{t \to \infty} {\cal X}(k,t), \mbox{etc}\\
\end{split}
\label{Notations}
\eeq
Extension to other quantities should be evident.

Let us begin with the time derivative of Eq.~(\ref{eq6.2}), 
which is given by
\beq \begin{split}
\partial_t G_{\th \r}(k,t) = & \Sigma_{\hth \hr}(k,t) {\cal X}(k,0) \\
& + \int_0^t ds \, \Sigma_{\hth \hr}(k,t-s) \partial_s {\cal X}(k,s) \\
& - \int_0^t ds \, [\partial_t \Sigma_{\hth \hth}(k,t-s)] \partial_s G_{\r\r}(k,s)
\end{split}
\label{eq6.44}
\eeq
Its long time limit is given by
\beq\begin{split}
0 = & ~\sigma_{\hth \hr}(k) {\cal X}(k,0)+\sigma_{\hth \hr}(k) 
\Big( x(k) -{\cal X}(k,0) \Big) \\
& - {\tilde \sigma}_{\hth \hth}(k) \Big( f_{\r\r}(k)- \r_0 S(k) \Big) \\
= & ~ \sigma_{\hth \hr}(k) x(k) 
 - {\tilde \sigma}_{\hth \hth}(k) \Big( f_{\r\r}(k)-\r_0 S(k) \Big) \ .
\end{split}
\label{eq6.45}
\eeq
Here ${\tilde \sigma}_{\hth \hth}(k) \equiv \lim_{t \to \io} 
[\partial_t \Sigma_{\hth \hth}(k,t)] $ vanishes since it  involves the time 
derivative of $G_{\r\r}(k,t)$ (see Eq.~(\ref{eq6.16})). 
Thus Eq.~(\ref{eq6.45}) reduces to 
\beq
\sigma_{\hth \hr}(k) x(k) = \sigma_{\hth \hr}(k) \Big(   K(k) f_{\r\r}(k)
+ f_{\th\r}(k) \Big) =0
\label{eq6.46}
\eeq

The one-loop expression of the self-energy 
$\Sigma_{\hth \hr}(k,t)=-\big( \Sigma_2 + \r_0 T k^2 \Sigma_{\hth\hth} \big)(k,t) $
 is easily obtained from Eq.~(\ref{eq6.16}) and Eq.~(\ref{eq6.19}) as
\beq \begin{split}
& \Sigma_{\hth\hr}(k,t)  
= \frac{T k^2}{2\r_0^3} \int_{\bf q} G_{\r\r}(q,t) G_{\r\r}(|{\bf k}-{\bf q}|,t) \\
& + \frac{T}{\rho^2_0} \int_{\bf q}  {\bf k}\cdot {\bf q}\, 
\Big[ \frac{U(q)}{T} 
G_{\rho\rho}(q,t) + G_{\th\rho}(q,t) \Big] G_{\rho\rho}(|{\bf k}-{\bf q}|,t) \\
& = \frac{T}{\rho^2_0} \int_{\bf q}  {\bf k}\cdot {\bf q}\, 
\Big[ K(q)
G_{\rho\rho}(q,t) + G_{\th\rho}(q,t) \Big] G_{\rho\rho}(|{\bf k}-{\bf q}|,t) \\
\end{split} 
\label{eq6.47}
\eeq
where we used $K(q) = U(q)/T+1/\r_0$.
Taking $t \to \io$ in (\ref{eq6.47}) gives
 \beq 
 \sigma_{\hth\hr}(k)  
= \frac{T}{\rho^2_0} \int_{\bf q}  {\bf k}\cdot {\bf q}\, 
\Big[ K(q) 
f_{\rho\rho}(q) + f_{\th\rho}(q) \Big] f_{\rho\rho}(|{\bf k}-{\bf q}|) .
\label{eq6.47a}
\eeq
Therefore, Eq.~(\ref{eq6.46}) and Eq.~(\ref{eq6.47a}) are incompatible with and thereby rule out the following nonergodic behaviors\\ 
(a) both $f_{\r\r}(k) \neq 0$ and $f_{\th\r}(k) \neq 0$, and \\
(b) both $f_{\r\r}(k) \neq 0$ and $f_{\th\r}(k) = 0$, \\
leaving out
the other two possibilities: \\
(c) both $f_{\r\r}(k) =0$ and $f_{\th\r}(k) \neq 0$, and \\
(d) both $f_{\r\r}(k) = 0$ and $f_{\th\r}(k) = 0$.
We now show that the nonergodic behavior (c) is not supported by the dynamics.
Taking the long-time limit of Eq.~(\ref{eq6.11}), we obtain
\beq \begin{split}
& f_{\th\r}(k) = \Big(\frac{1}{\r_0 S(k)}-K(k) \Big) f_{\r\r}(k) \\
& -\Big( n(k)+ \sigma_{\hth\hth}  (k) \Big) 
\Big( f_{\r\r}(k)-\r_0 S(k) \Big)
\label{eq6.48}
\end{split} \eeq
where $n(k)\equiv \lim_{t \to \io} {\cal N}(k,t) $.
On the other hand, the equation of motion for ${\cal N}(k,t)$ is easily 
obtained from Eq.~(40) as
\beq \begin{split}
\r_0 T k^2 \partial_t {\cal N}(k,t) = & ~ \partial_t \Sigma_{\hth\hr}(k,t) \\
& -\int_0^t ds \, \Sigma_{\hth\hr} (k, t-s) \partial_s \Sigma_{\hth\hr}(k,s) \\
& + \int_0^t ds \, \Sigma_{\hr\hr}(k,t-s) \partial_s {\cal N}(k,s) \ .
\label{eq6.49}
\end{split} \eeq
where $\Sigma_{\hth\hr}(k,0)=0$ and ${\cal N}(k,0)=0$ are used.
The long-time limit of this equation is then given by
\beq
0=  -(\sigma_{\hth\hr}(k))^2  +\sigma_{\hr\hr} (k) n(k)
\label{eq6.50}
\eeq

Now substituting $f_{\r\r}(k)=0$ into both Eq.~(\ref{eq6.48}) and Eq.~(\ref{eq6.50}), 
we respectively get
\beq 
f_{\th\r}(k)=\r_0 S(k) n(k)
\label{eq6.51}
\eeq
and 
\beq 
0=\sigma_{\hr\hr} (k) n(k).
\label{eq6.52}
\eeq
The former equation follows since 
$\sigma_{\hth\hth}(k) =0 $ for $f_{\r\r}(k)=0$ from (\ref{eq6.16}).
The latter one follows 
since  $\sigma_{\hth\hr}(k)$ is vanishing when $f_{\r\r}(k)=0$.

If the possibility (c) were to hold, then we would have 
\beq \begin{split}
& \sigma_{\hr\hr}(k)
=\sigma_1(k) + \r_0 T k^2 \sigma_{\hth\hr}(k) \\
& = 
\sigma_1(k) = T^2 \int_{\bf q} \, 
{\bf k} \cdot {\bf q} \,\, {\bf k} \cdot ({\bf k}-{\bf q}) f_{\th\r}(k) f_{\r\th}(k) \neq 0 
\end{split}
\label{eq6.53}
\eeq
where the last line results from the long-time limit of Eq. (\ref{eq6.22}) with 
$f_{\r\r}(k)=0$.
The equation (\ref{eq6.52}) and (\ref{eq6.53}) would then imply that
$n(k)=0$, which in turn would mean $f_{\th\r}(k)=0$ from (\ref{eq6.51}), which
contradicts the initial assumption $f_{\th\r}(k)\neq 0$.
Thus the possibility (c) cannot simultaneously satisfy Eqs.~(\ref{eq6.51}) and (\ref{eq6.52}). 
We are thus only left with
the ergodic behavior $f_{\r\r}(k)=0$ and $f_{\th\r}(k)=0$, which is 
the only possibility compatible with both Eq.~(\ref{eq6.51}) and Eq.~(\ref{eq6.52}).

Thus we find that the ENE phase transition
predicted by the  SMCT is absent when the full coupled dynamic equations with
the one-loop self-energies are considered; 
the equilibrium dynamics within one-loop theory always remains ergodic.
This dynamic aspect is a very much desired result 
since the FDR-preserving one-loop theory presented here
incorporates the SMCT, but at the same time goes beyond the SMCT, leading to
the ergodic equilibrium dynamics. 
In order to see how the full dynamic equations round off the sharp transition of the SMCT, 
 one needs to perform numerical integration of the full coupled dynamic equations
 with one-loop self-energies.

\section{Discussion and conclusion}
\label{conclusion}

\subsection{Discussion}

In this work, we reanalyzed the equilibrium dynamics of the DK equation for colloids 
via the FDR-preserving loop expansion for the new dynamic action 
incorporating an extra set of auxiliary fields $\{\th, \hth\} $, 
which is invariant under the linear  field transformations. 
In doing so, we rectified the incorrect point made in [KK08].
The authors missed a delta function contribution to the response function
$G_{\th\hth}(k,t)$, which led to the false linear dependence of 
$G_{\th \r}(k,t)$ and $G_{\th \th}(k,t)$ on $G_{\r\r}(k,t)$, and 
equivalently to the {\it invalid} FDR $ G_{\r\hr}(k,t)=i \Theta(t) G_{\r\r}(k,t)/\r_0 S(k)$.
These  invalid linear relations led the authors to 
obtain a nonperturbative closed equation for $G_{\r\r}(k,t)$ alone  in [KK08].
When the aforementioned missing contribution is reinstated,  
the full dynamic equations do not allow these simple relations between the correlation 
functions, and in turn do not lead to a single closed equation for $G_{\r\r}(k,t)$.
Instead, one really has to confront a  coupled set of dynamic equations for the correlation functions $G_{\r\r}(k,t) $, $G_{\th \r}(k,t) $, 
and $G_{\th \th}(k,t) $, which is ultimately subject to numerical integrations. 

In the present work, we elucidated how one can retrieve the SMCT
equation for $G_{\r\r}(k,t)$ from the full set of dynamic equations with the one-loop self-energies. 
In this process, we witnessed an amazing cancellation of the bare interaction potential, 
leaving  the resulting equation involving the direct correlation function only, which results from
a consistency between statics and dynamics. 
We also find that the SMCT is obtained in a much simpler fashion if from the outset  the  coarse-graining viewpoint is taken using the 
Ramakrishnan-Yousouff form of the free energy, i.e., Eq.~(\ref{eq2.3}) with $U(r) =-Tc(r)$.

The most important recognition in the present study is that 
the one-loop theory is not tantamount to the SMCT, and goes beyond
the SMCT; while the SMCT can be retrieved from the full dynamic equations
with a well defined approximation (and hence 
systematic corrections can be given) within the one-loop theory, 
the full one-loop  dynamic equations themselves are found to remove the sharp ENE transition 
predicted by the SMCT. 
How exactly the full one-loop dynamic equations smear out this sharp transition 
is a remaining key question, which is left for further study. 
It would be most fascinating to see, upon solving numerically the full one-loop dynamic equations, an initial SMCT-like dynamical regime, followed by a relaxation towards equilibrium. 

In the formulation of Jacquin and van Wijland \cite{JvW11}, the same situation was found, i.e. that one-loop theory was not 
enough to obtain a closed equation on the density sector,
and further simplifications (one-loop consistent substitutions) were needed, that led to the usual SMCT equation. Even more striking, upon application of this method to a very simple kinetically constrained model, 
which is not supposed to present an ENE transition of the MCT type, a SMCT-like equation was again recovered that led to a spurious ENE transition \cite{hugo-thesis}. However in this model it is easy to see, with the same line of 
thoughts as that followed in the previous subsection, that the transition 
cannot exist in the full theory without projection on the density sector, 
in accordance with independent \mbox{calculations \cite{WBG05}.}

Another line of work has been followed in order to incorporate momentum currents in the formalism, 
a process which was postulated to be at the origin of the cut-off of the ENE
transition \cite{DM86,GS87}, and later challenged \cite{CR06}.  
 In the present field theory setting extended to non-linear fluctuating hydrodynamics, the situation has been instead 
shown \cite{NH08,DM09,yeo} to be similar to the simple DK case, and thus 
we expect our conclusions to hold also in that case, even though a direct proof would be of course quite cumbersome. 

The present findings seem to pose a challenge to our understanding on the nature of the SMCT. The DK equation describes the relaxation dynamics towards the well-defined equilibrium state, and therefore it is natural that the ergodic-nonergodic phase transition is absent in the DK equation.
Then it would be interesting to ask in what approximation or in which limit one can get the SMCT. 
It would be interesting to show that the SMCT can be recovered from a more physically grounded limit such as 
$N \to \infty$ in a $N$-component generalized DK equation.

The SMCT does not seem to correspond to a usual simple mean field theory, but rather to an unusual one with subtle aspects.
It would thus be desirable to investigate the high dimensional version of the present set of equations, in order to see whether 
coherent results are obtained in this limit, which is usually associated to mean-field, since in infinite dimensions, every particle has an infinite number of nearest neighbors.

The SMCT equation in high dimensions was recently studied \cite{schmid,ikeda_miyazaki,CIPZ11,CIPZ12},  and the equation was shown in \cite{ikeda_miyazaki,CIPZ11} to break down in this limit.
Whether this is still the case 
in the hereby proposed dynamical equations remains to be investigated. 

A major concern with the present formulation is that it is an expansion around a Gaussian theory, 
and not around the interaction-free case. 
Thus at each loop order, one is likely to break the exact evolution equation of the density in the NBG.
For example one can see that the one-loop self energies do not cancel straightforwardly when evaluated at $U=0$, and 
even though the solution of the full set of equations with $U=0$ can well be given by the NBG result, it is not guaranteed 
by the formalism (although the full action treated exactly of course correctly treats the $U=0$ limit, as was already shown in [KK08]).

The NBG case is a strongly non-Gaussian theory, as one can already see at the static level, where all the cumulants of the density, and 
their associated vertex functionals are non-zero in the NBG. This has also been checked in the dynamical framework in \cite{VCCK08}.
Thus nonperturbative relations like the first line of Eq.~(\ref{eq3.13})  
(required by the consistency between statics and dynamics) are likely to hold, and to be broken in a naive loop expansion as usually performed. 
As an explicit example, for $U=0$ case, the first line of Eq.~(\ref{eq3.13}) is not fulfilled by Eq.(43), the one-loop expression for $\S_{\hth\hth}(k,t)$.
As we indicated in the introduction, a possible remedy would be to directly work with the conjugate fields $\{ \r,\hr\}$, and expand around the non-Gaussian 
NBG theory, utilizing the T-transformation.
Another type of approach, chosen in \cite{Sz07} and \cite{Ma11}, have the potential to overcome
these difficulties, but due to the specificities of the theories developed in these works, progress is likely to be difficult.

\subsection{Conclusion}

We have amended a previous work aiming at rederiving the SMCT equation from a field-theoretic treatment of the
Brownian dynamics of an equilibrium glass. We found that the SMCT equation is not obtained from a simple one-loop self-consistent approximation, but rather from a further step of approximation, that we identified in a clear way.
We have shown that taking from the outset
 a coarse-grained point of view by replacing $U$ by $-Tc(r)$ leads back again to the SMCT equation, thus reconciling our approach with previous works \cite{KK94} based on this coarse-grained point of view.
We have shown that the full one-loop dynamical equations does not support the existence of an ENE transition, which is not without reminding the results obtained in \cite{ABL06}, and that the forceful projection of the dynamics on the density-density sector leads to this spurious ENE transition, in accordance with the results obtained in \cite{JvW11}.

However, we emphasize the fact that the present loop expansion, as opposed to that in \cite{JvW11}, is not an expansion 
around the NBG limit, and is thus likely to produce
spurious results if taken as is. 
A deeper understanding of the high-order sum rules constraining the theory should be obtained 
before further analyses of our set of equations.

\acknowledgments

We gratefully acknowledge discussions with A. Andreanov, G. Biroli, S. Das, M. Fuchs, H. Hayakawa, J. Horbach, H. L\"owen, G. F. Mazenko, K. Miyazaki,  A. Onuki, S. Sasa, G. Szamel, J. Yeo, and H. Yoshino. 
B.K. was supported by Basic Science Research Program through
the National Research Foundation (NRF) funded by the Ministry of Education, Science, 
and Technology (Grant No. 2011-0009510), and in part by
Changwon National University grant (2011-2013).
H.J. was supported by a CFM-JP Aguilar grant when this work was initiated.

\begin{appendix}

\section{Derivation of Eqs.~(\ref{eq6.9}) and (\ref{equationforkernelN})}
\label{app_derivation-kernels}

We first define the Laplace transform
\beq
A^L(z) \equiv {\cal L}(A)(z)= \int_0^{\infty} dt \, e^{-zt} A(t)
\label{e1}
\eeq
and the function $a(k)$ as:
\beq
a(k)=\r_0 T k^2
\eeq
Dropping the wave-vector dependance, we rewrite (\ref{eq6.1}) as 
\beq\begin{split}
& a {\cal X}(t) = -\dot G_{\rho\rho}(t) \\
& + \int_0^t  ds\, \Big[ \Sigma_{\hat\rho \hat\rho}(t-s) {\cal X}(s) - \Sigma_{\hat\rho \hat\theta}(t-s)   \dot G_{\rho \rho}(s) \Big] \ .
\end{split}
\label{e2}
\eeq
Its Laplace transform is given by
\beq
a \Big( 1-  \frac{1}{a} \Sigma_{\hat\rho \hat\rho}^L(z)  \Big) {\cal X}^L(z)
=-\Big( 1 +  \Sigma_{\hat\rho \hat\theta}^L(z) \Big) {\cal L}(\dot G_{\r\r})(z)
 \label{e3}
\eeq

A new memory function ${\cal M}(t) $ was defined in (\ref{eq6.8}) as
\beq
\int_0^t  ds\, \Sigma_1  (t-s) {\cal X}(s)
\equiv 
-\int_0^t  ds\, {\cal M}(t-s) \dot G_{\rho\rho}(s)
\label{e4}
\eeq
Its Laplace transform is given by
\beq
\Sigma_1^L(z)  {\cal X}^L(z) =-{\cal M}^L(z) {\cal L}(\dot G_{\rho\rho})(z)
\label{e5}
\eeq
Multiplying both sides of (\ref{e3}) by $a^{-1}\Sigma_1^L(z)$ 
and using (\ref{e5}), 
we obtain
\beq
\Big( 1-  \frac{1}{a} \Sigma_{\hat\rho \hat\rho}^L(z) \Big) {\cal M}^L(z)
=\frac{1}{a} \Big( 1 +  \Sigma_{\hat\rho \hat\theta}^L(z) \Big)
\,  \Sigma_1^L (z)
\label{e6}
\eeq
Rearranging this equation gives
\beq
{\cal M}^L(z) =  \frac{1}{a} \Big[  \Sigma_1^L (z)  
+
 \Sigma_{\hat\rho \hat\theta}^L(z) \Sigma_1^L (z) 
+
 \Sigma_{\hat\rho \hat\rho}^L(z)  {\cal M}^L(z) \Big].
\label{e7}
\eeq
Putting this equation back in the time domain, we get
\beq \begin{split}
a {\cal M}(t) = & ~ \Sigma_1(t) - \int_0^t ds \, 
 \Sigma_1(t-s)  \Sigma_{ \hat\theta \hat\rho }(s) \\
& + \int_0^t ds \,  \Sigma_{\hat\rho \hat\rho}(t-s) {\cal M}(s)
\end{split}
\label{e8}
\eeq
This is Eq.~(\ref{eq6.9}) when the wavenumber is restored.

Another memory function ${\cal N}(t) $ was defined in (\ref{defkernelN}) as
\beq
\int_0^t  ds\, \Sigma_{\hth\hr}  (t-s) {\cal X}(s)
\equiv 
-\int_0^t  ds\, {\cal N}(t-s) \dot G_{\rho\rho}(s).
\label{e9}
\eeq
Its Laplace transform is given by
\beq
\Sigma_{\hth\hr}^L(z)  {\cal X}^L(z) =-{\cal N}^L(z) {\cal L}(\dot G_{\rho\rho})(z).
\label{e10}
\eeq
Let us multiply both sides of (\ref{e3}) by $a^{-1}\Sigma_{\hth\hr}^L(z)$.
Using (\ref{e10}), we get
\beq
\Big( 1-  \frac{1}{a} \Sigma_{\hat\rho \hat\rho}^L(z) \Big) {\cal N}^L(z)
=\frac{1}{a} \Big( 1 +  \Sigma_{\hat\rho \hat\theta}^L(z) \Big)
\,  \Sigma_{\hth\hr}^L (z).
\label{e11}
\eeq
Rearranging this equation gives
\beq
{\cal N}^L(z) =  \frac{1}{a} \Big[   \Sigma_{\hth\hr}^L (z)  
+
 \Sigma_{\hat\rho \hat\theta}^L(z) \Sigma_{\hth\hr}^L (z) 
+
  \Sigma_{\hat\rho \hat\rho}^L(z)  {\cal N}^L(z) \Big].
\label{e12}
\eeq
This is transformed back in the time domain as:
\beq 
\begin{split}
a {\cal N}(t) = & ~ \Sigma_{\hth\hr}(t) - \int_0^t ds \,  
\Sigma_{\hth\hr}(t-s)  \Sigma_{\hth\hr}(s) \\
& + \int_0^t ds \,  \Sigma_{\hat\rho \hat\rho}(t-s) {\cal N}(s).
\end{split}
\label{e13}
\eeq
where the identity $\Sigma_{\hr \hth}(t)=-\Sigma_{\hth \hr}(t) $ is used.
This is (\ref{equationforkernelN}) when the wavenumber is restored.

\end{appendix}

\end{document}